\documentclass[preprint, authoryear, 12p]{elsarticle}
\usepackage{amsmath}
\usepackage{amssymb}
\usepackage{graphicx}

\makeatletter
\usepackage{hyperref}

\journal{Planetary and Space Science}

\begin{document}

\begin{frontmatter}

\title{Toward a numerical deshaker for PFS}

\author{Schmidt$^{1,2}$ F., Shatalina$^{3}$ I., Kowalski$^{4}$ M., Gac$^{4}$
N., Saggin$^{3}$ B., Giuranna$^{5}$, M.}

\address{
$^{1}$Univ. Paris-Sud, Laboratoire IDES, UMR 8148, Bât 509, Orsay,
F-91405, France (frederic.schmidt@u-psud.fr)

$^{2}$ CNRS, Orsay, F-91405, France

$^{3}$ Dipartimento di Meccanica, Politecnico di Milano, Campus of
Lecco, Via M. d\textquoteright{}Oggiono 18/a, 23900, Lecco, Italy

$^{4}$ Laboratoire des signaux et systemes (L2S), UMR8506 Univ Paris-Sud
- CNRS \textendash{} SUPELEC, SUPELEC, 3 rue Joliot Curie, Gif sur
Yvette, F-91192, France

$^{5}$ IFSI, via del Fosso del Cavaliere, 100, 00133 Roma, Italy}

\begin{abstract}

The Planetary Fourier Spectrometer (PFS) onboard Mars Express (MEx)
is the instrument with the highest spectral resolution observing Mars
from orbit since January 2004. It permits studying the atmospheric
structure, major and minor compounds. The present time version of
the calibration is limited by the effects of mechanical vibration,
currently not corrected. We proposed here a new approach to correct
for the vibrations based on semi-blind deconvolution of the measurements.
This new approach shows that a correction can be done efficiently
with 85\% reduction of the artifacts, in a equivalent manner to the
stacking of 10 spectra. Our strategy is not fully automatic due to
the dependence on some regularisation parameters. It may be applied
on the complete PFS dataset, correcting the large-scale perturbation
due to microvibrations for each spectrum independently. This approach
is validated on actual PFS data of Short Wavelength Channel (SWC),
perturbed by microvibrations. A coherence check can be performed and
also validate our approach. Unfortunately, the coherence check can
be done only on the first 310 orbits of MEx only, until the laser
line has been switch off. More generally, this work may apply to numerically
\textquotedblleft{}deshake\textquotedblright{} Fourier Transform Spectrometer
(FTS), widely used in space experiments or in the laboratory.
\end{abstract}


\end{frontmatter}

\section{Introduction}

The Planetary Fourier Spectrometer (PFS) is a double pendulum Fourier
transform infrared spectrometer instrument onboard MEx, operating
in the 1.2 to 5.5 micrometers for the Short Wavelength Channel (SWC),
and 5 to 45 microns in the Long Wavelength Channel (LWC) \citep{Formisano_PFS_PSS2005}.
It is based on a modified Michelson's scheme using a double pendulum
with cubic reflectors. The optical path difference is defined by the
zero crossing of a laser tacking the same optical path asthe signal.
The spectra presented in this article are the numerical Fourier transform
of the recorded interferograms.

An experimental study of mechanical vibration impact on Fourier-transform
spectrometer has been proposed based on PFS example \citep{Comolli_Evaluationsensitivityto_RSI2005}.
Analytical expression of all distortion effects have been formulated
separately \citep{Saggin_PFSvibration_AO2007}: offset of the reference
laser signal, mirrors speed variation, periodic misalignments, detector
non linearity and internal reflections. More recently, a numerical
simulation model has been proposed to explore all effects combined
in order to understand the PFS signal \citep{Comolli_ProbPFS_PSS2010}.
Perturbations are creating artificial features, called ``ghosts'',
present in some spectra of the SWC but not in the LWC, thanks to the
optimization of the pendulum velocity \citet{Giuranna_calibrationPFS-SWC_PSS2005,Giuranna_calibrationPFS-LWC_PSS2005}.
Since the amplitude of ghosts are small (few \% of the original signal)
and its phase has a stochastic behavior, the worst cases correspond
to only few significant ghosts \citet{Shatalina_PFS_AOsubmitted}. 

Quantitatively, the ghosts are affecting few \% of the total spectrum
energy (3\% typically; 5\% maximum). When single spectra are used,
the absolute radiometric calibration is degraded, and spurious spectral
features may appear in the spectrum, preventing any surface-related
analysis, and introducing possible large uncertainties in the quantitative
retrievals of abundances of minor species in the atmosphere. When
discussing the calibration procedure for the SWC \citep{Giuranna_calibrationPFS-SWC_PSS2005}
and the LWC \citep{Giuranna_calibrationPFS-LWC_PSS2005}, the authors
suggest to stack the data to correct for the effects of the mechanical
vibrations. The position of ghosts depends on the frequencies of the
external vibrations, which have been found to be quite stable. Since
the phase of ghosts is random and the external frequencies are stable,
only the signal should be coherent during the stack. This idea has
been confirmed by numerical modeling of the perturbations \citep{Comolli_ProbPFS_PSS2010}.
Practically, averaging a few spectra (ten or so) is enough to average
out the ghosts. However, this will degrade the spatial and temporal
resolution of PFS measurements, limiting the interpretation of small-scale
features and hampering some scientific studies (e.g., the composition
of ices; detection of minerals at the surface).

\begin{figure}
\includegraphics[clip,width=1\columnwidth]{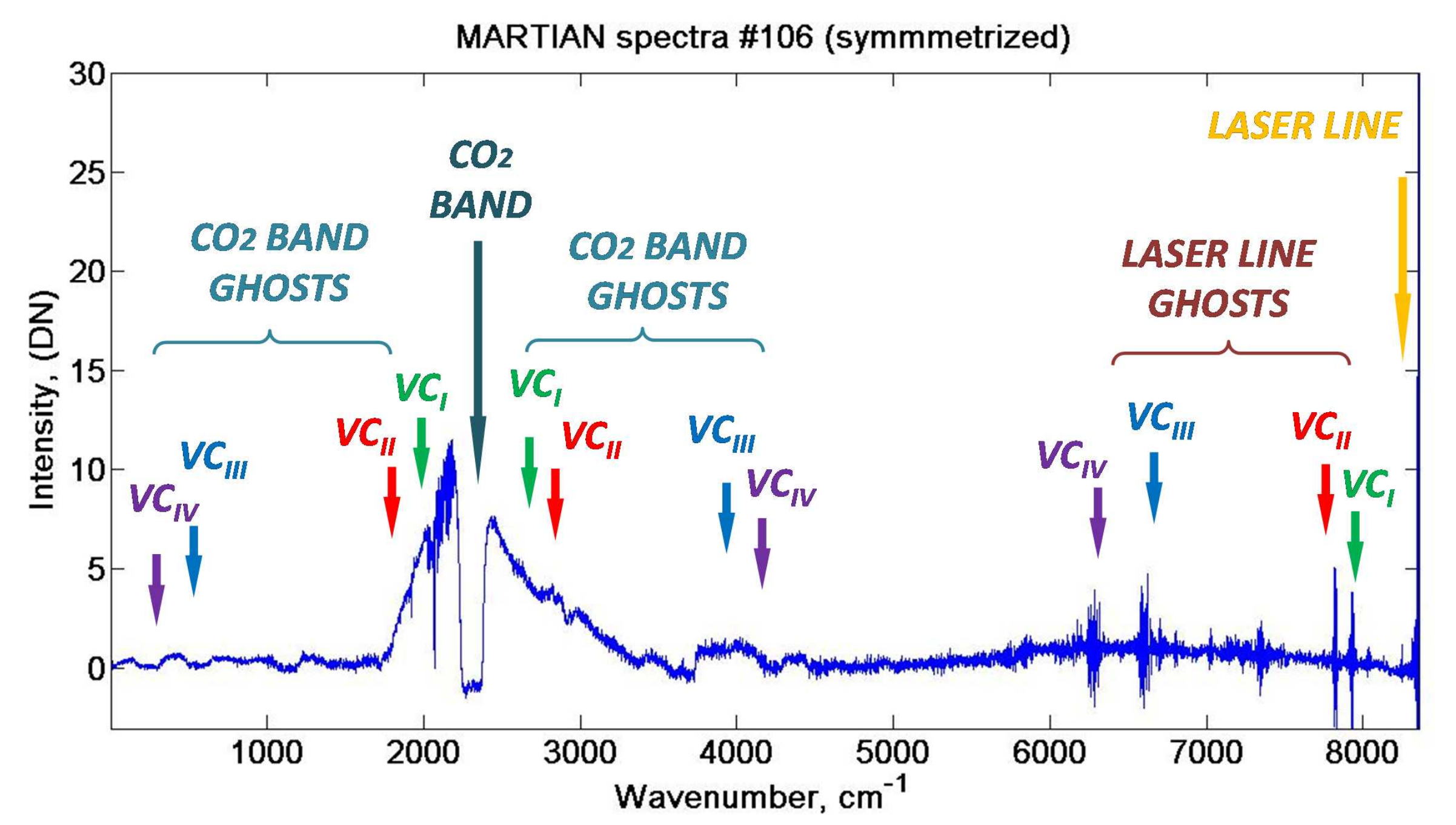}\center

\caption{Typical symmetrized PFS measurement in SWC. Signal and major CO$_{2}$
band and laser lines are noted. The four main ghosts are identified
as ``Vibration Component'' (VC) affecting both signal and laser
line.. \label{fig:TypicalPFS}}
\end{figure}

Typical PFS raw measurements are shown in Fig. \ref{fig:TypicalPFS}.
One can identify the major signals from Mars: thermal emission and
reflection of solar energy, and the laser line stray-light. Also the
contribution due to mechanical vibrations are shown on the signal,
leading to additional energy shifted on left and right almost symmetrically.
The ghost of the laser line is only one sided due to aliasing.

Our aim is to provide a new approach to process the PFS instrument
with following constraint: 
\begin{enumerate}
\item correct the effect of mechanical vibrations due to both misalignment
and optical path difference errors.
\item perform the correction on each spectrum separately.
\item validate the approach by using actual PFS observations.
\end{enumerate}

In order to avoid unphysical solution, the algorithm is initialized
with an a priori guess of the large scale structure of the spectra,
adapted to each measurement, reproducing the Martian thermal emission
and the reflected solar light (see section \ref{sub:Initial-guesses}).

A check of the correction can be done but requires the SWC laser diode
switched on, to estimate the vibration kernel independently (see section
\ref{sub:Laser-line}).

\section{Method}

This section describes the direct model of the Martian spectra affected
by vibrations. Then, an iterative procedure is exposed in order to
invert it as well as some criteria to measure the quality of our estimation.

\subsection{Analytical formulation in the signal domain (0 to 5000 cm$^{-1}$)}

As it can be seen from Fig.\ref{fig:TypicalPFS}, it is possible to
separate the whole spectrum into two wavenumber domains to deal the
effects of the mechanical vibrations apart in each of them. From 0
to 5000 points (5000{*}1.02 cm$^{-1}$), we define the signal domain,
where the thermal energy from Mars and the most of the reflected Martian
energy are recorded, without significant laser line artefacts. The
laser line domain is defined from 5000 cm$^{-1}$ to 8330 cm$^{-1}$.
It contains also the Martian signal but affected by laser line artefacts. Below
1700 cm$^{-1}$ there is no meaningful signal due to the low detector
responsivity (see Fig. 15. in \citet{Giuranna_calibrationPFS-SWC_PSS2005}),
and this region is characterized only by ghosts of the continuum.

At larger wavenumber than 5000 cm$^{-1}$, the signal is affected
by the laser line shape and its ghosts, directly and in aliasing.
This domain will be used to test the coherence of the results. Since
the laser has been switched off after orbit 634, it could not be used
for the complete PFS archive (see section \ref{sub:Laser-line}).

Using some mathematical reorganization and simplification, the analytical
expression of mechanical vibration due to periodic misalignment and
optical path errors can be written as a convolution products in complex
form, see Eq. (13) in \citet{Shatalina_PFS_AOsubmitted}. Assuming
that the domain of wavenumber with significant signal $I_{Mars}$
around $\sigma\thicksim2000-3000$ cm$^{-1}$ is constant ($\sigma_{k}\thicksim2500$),
the following Equation:
\begin{equation}
I_{PFS}(\sigma)=I_{Mars}(\sigma)+\left[\sigma.I_{Mars}(\sigma)\right]\star K(\sigma)\ ,
\end{equation}
simplifies to: 

\begin{equation}
I_{PFS}(\sigma)=I_{Mars}(\sigma)\star\left[\delta(\sigma)+K(\sigma).\sigma_{k}\right]\ .
\end{equation}
with $\delta(),$the dirac function.

By rewriting:

\begin{equation}
I_{PFS}(\sigma)=I_{Mars}(\sigma)\star K_{PFS}(\sigma)\ ,
\end{equation}
with $I_{PFS}(\sigma)$ the measured raw spectra, $I_{Mars}$ the
contribution of the raw spectra from Mars, $K_{PFS}$ the kernel representing
the mechanical vibration effects, $\sigma$ the wavenumber, and $K(\sigma)$
the non-normalized complex kernel \citep{Shatalina_PFS_AOsubmitted}.

From \citet{Shatalina_PFS_AOsubmitted}, the kernel of all frequency
of vibrations is:

\begin{equation}
K_{PFS}(\sigma)=\delta(\sigma)+A(\sigma)e^{i\varphi_{A}(\sigma)}+B(\sigma)e^{i\varphi_{B}(\sigma)}\ .
\end{equation}
The quantities A, B, $\varphi_{A}$, $\varphi_{B}$ are unknown and
cannot be evaluated quantitatively due to the lack of knowledge about
vibration amplitude and phase. In practice, the functions A, B, $\varphi_{A}$,
$\varphi_{B}$ are sparse over $\sigma$ because the frequencies of
vibrations are sparse. Please note that A, B, $\varphi_{A}$, $\varphi_{B}$
are not symmetric around $\sigma=0$ due to the relative phase. We
propose to estimate those functions using an inversion procedure described
in the next section.

The assumption of a reduced wavenumber domain is valid in first approximation
due to the sensitivity of the detector and the typical Martian signal,
leading to a misfit factor of x0.8 to x1.2 that is reasonable for
this case. In addition, our strategy is to use semi-blind deconvolution
algorithm in order to ensure the best fit any kind of spectra. This
way, the wavenumber domain of significant signal has not to be defined
explicitly.

\begin{figure}
\includegraphics[width=1\textwidth]{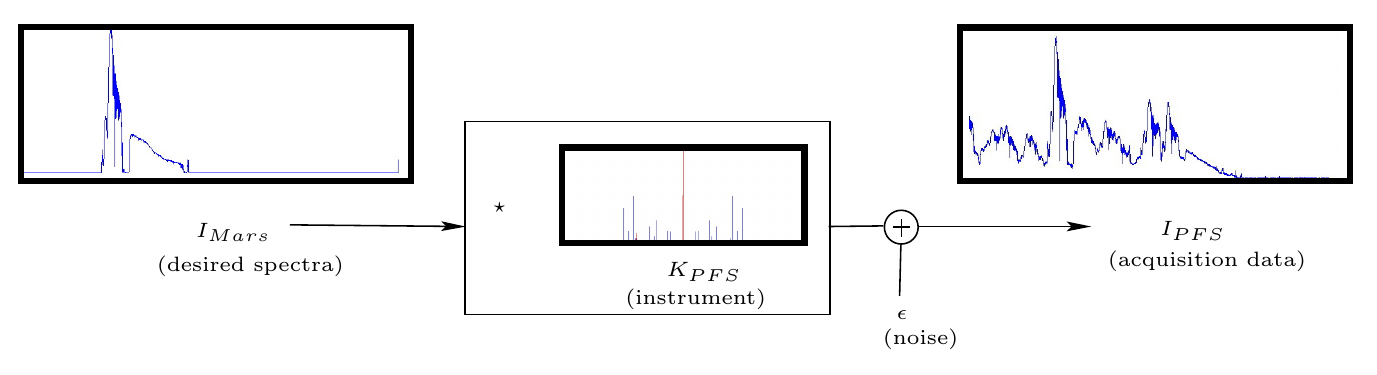}\centering

\caption{Model of acquisition by the PFS instrument \label{fig:Modelisation-of-acquisition}}
\end{figure}

Including to our model an additive noise $\epsilon$ which stands
for the others sources of acquisition noise besides the mechanical
vibrations and the error due to our PFS modeling by a convolution
kernel $K_{PFS}$, PFS spectra in signal domain as illustrated in
Fig. \ref{fig:Modelisation-of-acquisition}, are obtained through:

\begin{equation}
I_{PFS}(\sigma)=I_{Mars}(\sigma)\star K_{PFS}(\sigma)+\epsilon\ .\label{eq:direct-model}
\end{equation}

\subsection{Inversion}

From the direct model of the PFS instrument described above, see Eq.
(\ref{eq:direct-model}), we propose here a semi-blind deconvolution
method to solve the inverse problem: estimation of the desired spectra
$I_{Mars}$ from the PFS spectra $I_{PFS}$ although the convolution
kernel $K_{PFS}$ is unknown. We qualified our method as semi-blind
because the only spectral a priori information is $\hat{I}_{Mars}^{0}$,
known ab initio. We also used two a priori information : $\hat{I}_{Mars}$
is smooth and $\hat{K}_{PFS}$ is sparse. The notation $\hat{X},$means
the estimation of quantity $X$. A classical approach consists in
introducing a cost function $\mathcal{C}$ whose minimum provides
an estimation:

\global\long\def\argmin{\mathop{\mathrm{argmin}}}
 
\begin{align}
\hat{I}_{Mars},\hat{K}_{PFS}= & \argmin_{I_{Mars},K_{PFS}}\mathcal{C}(I_{Mars},K_{PFS})\nonumber \\
= & \argmin_{I_{Mars},K_{PFS}}\frac{{1}}{2}\|I_{PFS}-K_{PFS}\star I_{Mars}\|_{2}^{2}+\lambda_{K}\|K_{PFS}\|_{1}+\frac{{\lambda_{Mars}}}{2}\|D\star I_{Mars}\|_{2}^{2}\ .\label{eq:functional}
\end{align}

Three terms appear in $\mathcal{C}$
\begin{enumerate}
\item A data fit term $\frac{{1}}{2}\|I_{PFS}-K_{PFS}\star I_{Mars}\|_{2}^{2}$
that quantifies how well the estimated sources match the measured
data. This term takes into account the characteristics of the noise
supposed to be white and gaussian. This data match term is sensitive
to high frequency noise and must be balanced with regularization term
which corresponds to a mathematical prior on the expected solution
\citep{Idier08a}.
\item A sparsity regularization term $\|K_{PFS}\|_{1}$ is chosen for the
kernel, i.e. : the $\ell_{1}$ norm (sum absolute value) of the kernel
must be low. Indeed, the PFS kernel is supposed to be composed with
few diracs at mechanical vibration frequencies.
\item A smooth regularization term is chosen for the Mars spectra : $\|D\star I_{Mars}\|_{2}^{2}$,
where $D$ is a discrete first-order derivation operator. This prior
promotes smooth solution in order to avoid noise improvement.
\end{enumerate}
All these terms are balanced with two hyperparameters $\lambda_{K}$
and $\lambda_{Mars}$ , both positive. The functional \ref{eq:functional}
is convex for each variables - convex in $I_{Mars}$ when $K_{PFS}$
is fixed and vice versa - but not from the couple $(I_{Mars},K_{PFS})$.
The strategy we choose here is a classical alternative procedure:
from initial guesses $\hat{I}_{Mars}^{0}$, an iterative procedure
updates successively at each iteration $n$, the new estimates $\hat{K}_{PFS}^{n+1}$
and $\hat{I}_{Mars}^{n+1}$.

\subsubsection{Iterative procedure }

At each iteration $n$ we estimate successively the kernel $\hat{K}_{PFS}^{n+1}$
and the signal $\hat{I}_{Mars}^{n+1}$ until iteration $N$ using
the following steps: 
\begin{enumerate}
\item First estimation of the kernel $\hat{K}_{PFS}^{1}$ from filtered
$\hat{I}_{Mars}^{0}$ and $I_{PFS}$ with L1 regularization
\item Iterative loop:

\begin{enumerate}
\item estimation of the Mars spectra $\hat{I}_{Mars}^{n+1}$ from unfiltered
$\hat{K}_{PFS}^{n}$ and $I_{PFS}$ with smooth regularization
\item estimation of the kernel $\hat{K}_{PFS}^{n+1}$ from unfiltered $\hat{I}_{Mars}^{n+1}$
and $I_{PFS}$ with L1 regularization
\end{enumerate}
\item Last estimation of the Mars spectra $\hat{I}_{Mars}^{final}$ from
unfiltered $\hat{K}_{PFS}^{final}$ and $I_{PFS}$
\end{enumerate}
For both estimations, a convex optimization algorithm converges to
the solution defined by the minimum of a criteria made of a data match
and a regularizations terms. This means that the solution is unique
and can be estimated either analytically or iteratively. 

Since the first step of the iterative procedure is the estimation
of the kernel $\hat{K}_{PFS}^{1}$, the only a priori information
of this iterative procedure $\hat{I}_{Mars}^{0}$, estimated ab initio.
Since $\hat{I}_{Mars}^{0}$, can only be estimated at large scale
(all absorption lines may differ from spectra to spectra due to non-homogeneity
of chemical compounds in the atmosphere/surface of Mars), the first
iteration is done in a low-pass filtered space, as described in section
\ref{sub:Initial-guesses}.

\paragraph{Estimation of the PFS kernel}

The estimation of the PFS kernel reduce to the following $\ell_{1}$
regularized convex (non smooth) problem:

\begin{equation}
\hat{K}_{PFS}^{n+1}=\argmin_{K_{PFS}}\frac{{1}}{2}\|I_{PFS}-K_{PFS}\star\hat{I}_{Mars}^{n}\|_{2}^{2}+\lambda_{K}\|K_{PFS}\|_{1}\ ,\label{eq:pfs-kernel-estimation-1}
\end{equation}
where $\hat{I}_{Mars}^{n}$ is the estimation of the Mars spectra
at the iteration number $n$. This problem is the well known Lasso
\citep{Tibshirani96Lasso} or Basis-Pursuit Denoising \citep{Chen98atomic}
problem, and can be solved efficiently with the Fast Iterative/Thresholding
Algorithm (FISTA) \citep{Beck09Fista}. Denoting by $\tilde{I}$ the
adjoint of the kernel $I$ and by $S_{\lambda}$ the so-called soft-thresholding
operator%
\footnote{$S_{\lambda}(x)=\frac{{x}}{|x|}\max(|x|-\lambda,0)$%
} the algorithm reads:
\begin{enumerate}
\item Let $i=0,\:\tau^{0}=1,\: k=1,\ Z\text{\textsuperscript{0}}=K_{PFS}^{n}$
and $L=\|I_{Mars}\|^{2}$.
\item $K_{PFS}^{i}=S_{\lambda_{K}/L}\left(Z^{i}+\frac{1}{L}(I_{PFS}-Z^{i}\star\hat{I}_{Mars}^{n})\star\tilde{\hat{I}}_{Mars}^{n}\right)$
\item $\tau^{i+1}=\frac{{1+\sqrt{{1+4\tau^{i^{2}}}}}}{2}$
\item $Z^{i+1}=K_{PFS}^{i}+\frac{{\tau^{i}-1}}{\tau^{i+1}}(K_{PFS}^{i}-K_{PFS}^{i-1})$
\item $i=i+1$
\item Go to 2 until $i=i_{max}$
\item $K_{PFS}^{n+1}=K_{PFS}^{i_{max}}$
\end{enumerate}
From theoretical consideration, the kernel $K_{PFS}$ must be a dirac-shape
on zero, so we concentrate the energy around zero into a dirac to
create the kernel estimation $\hat{K}_{PFS}^{n}$. We would like to
emphasize that there is no analytical solution of eq. \ref{eq:pfs-kernel-estimation-1}
so we solve this equation with an iterative procedure, initialized
with the previous step $K_{PFS}^{n}$. For the first initialization
$K_{PFS}^{0}$, we may use $\hat{K}_{PFS}^{approx}$ but any other
guess (such zero) may apply when the laser line has been switch off.
Nevertheless, closer the initialization, faster the convergence.

\paragraph{Estimation of the Mars spectra}

For the Mars spectra, the estimation reduces to a classical Thikonov
regularization \citep{Idier08a}:

\begin{equation}
\hat{I}{}_{Mars}^{n+1}=\argmin_{I_{Mars}}\|I_{PFS}-\hat{K}_{PFS}^{n+1}\star I_{Mars}\|_{2}^{2}+\lambda_{Mars}\|D\star I_{Mars}\|_{2}^{2}\label{eq:mars-spectra-estimation-1}
\end{equation}
Thanks to the fact that a convolution is diagonal in the Fourier domain,
and the Parseval theorem, the solution reads:

\begin{equation}
\mathcal{F}(\hat{I}_{Mars}^{n+1})=\argmin_{\mathcal{F}(I_{Mars})}\|\mathcal{F}(I_{PFS})-\mathcal{F}(\hat{K}_{PFS}^{n+1})\odot\mathcal{F}(I_{Mars})\|_{2}^{2}+\lambda_{Mars}\|\mathcal{F}(D)\odot\mathcal{F}(I_{Mars})\|_{2}^{2}
\end{equation}

where $\odot$ is the Hadarmard element-wise product and $\mathcal{F}$
the Fourier transform. Then, the estimation of the Mars spectra at
iteration $n+1$ is given in close form by : 

\begin{equation}
\hat{I}_{Mars}^{n+1}=\mathcal{F}^{-1}\left(\mathcal{F}(I_{PFS})\odot(\mathcal{F}(\hat{K}_{PFS}^{n+1})^{-2}-\lambda_{Mars}\mathcal{F}(D)^{-2})\right)\label{eq:solution_mars_spectra_estimation}
\end{equation}

\[
\ ,
\]
where $\mathcal{F}(\hat{K}_{PFS}^{n+1})^{-2}$ (resp. $\mathcal{F}(D)^{-2}$
) represents the vector containing the inverted squared elements of
the vector $\mathcal{F}(\hat{K}_{PFS}^{n+1})$ (resp. $\mathcal{F}(D)$
). 

We would like to emphasize that equation \ref{eq:solution_mars_spectra_estimation}
is the analytical solution of eq \ref{eq:mars-spectra-estimation-1}
that did not require initialization.

\subsubsection{Initial guesses of the Martian spectra and PFS kernel \label{sub:Initial-guesses}}

\paragraph*{Initial Mars spectra guess}

We estimate the Martian spectra large scale feature (noted $\hat{I}_{Mars}^{0}$)
by two Planck functions and the major absorption feature, representing
(i) the Martian thermal emission and (ii) the solar energy reflected
back by Mars and (iii) the 2200-2400 cm$^{-1}$ gap, representing
the CO$_{2}$ absorption band. The Martian temperature is estimated
by fitting the 2500-3000 cm$^{-1}$ domain, where the ghost seems
to be less pronounced. The Planck function of the sun is scaled to
the 3800-4200 cm$^{-1}$ domain. We derive the raw spectra using the
calibrations of detector responsivity and deep space measurements
\citep{Giuranna_calibrationPFS-SWC_PSS2005}. This initial guess is
only valid at large scale because the absorption lines of major and
minor gases may change, due to local pressure, atmospheric circulation,
surface change and radiative transfer effects.

The phase of the initial guess is taken similar to the signal in the
domain where the ghosts are absent and a constant extrapolation is
proposed to the ghosted region.

Because the iterative procedure is sensitive to initialization, both
PFS spectra $I_{PFS}$ and mars initial guess $\hat{I}_{Mars}^{0}$
are filtered with a low-pass filter with\emph{ }a cut off frequency
of $\frac{1}{20\triangle\sigma}$, where $\triangle\sigma$ = 1.02
cm\emph{$^{-1}$} is the spectral resolution, in order to keep the
realistic features.

The initial guess $\hat{I}_{Mars}^{0}$ will force the initial step
of the iterative procedure to find a local minimum around physical
solution. Initializing the procedure with random or constant signal
lead to non physical solutions.

\begin{figure}
\includegraphics[clip,width=1\textwidth]{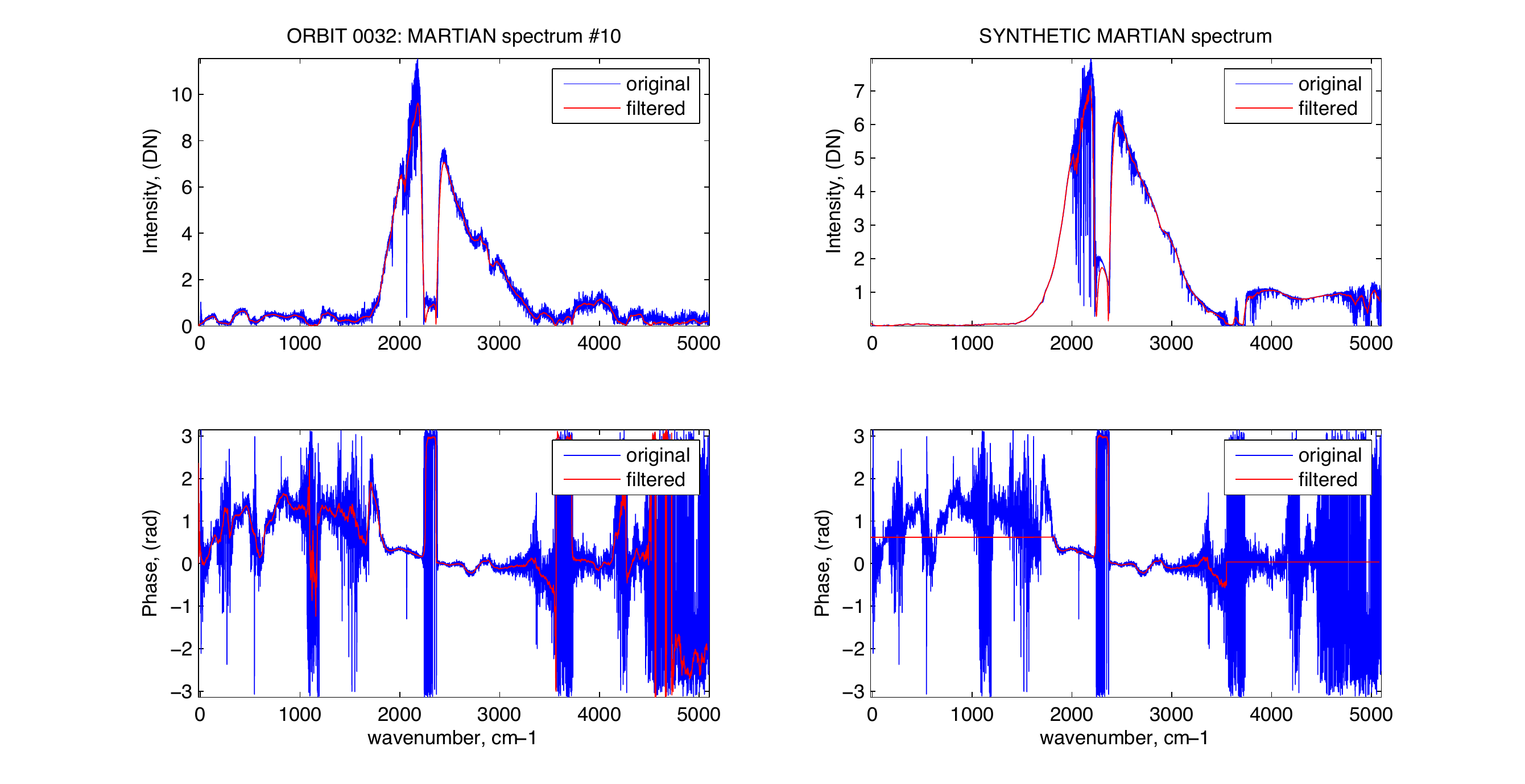}\center

\caption{Raw measurements $I_{PFS}$ (on left) and initial guess of the Martian
signal $\hat{I}_{Mars}^{0}$ (on right) for the PFS measurement ORB0032,
No 106. \label{fig:GuessSignal}}
\end{figure}

\paragraph*{Initial PFS Kernel guess}

Thanks to the approximation from \citealp{Shatalina_PFS_AOsubmitted},
an estimation of the kernel $\hat{K}_{PFS}^{approx}$ from the laser
line domain can be done (see section \ref{sub:Laser-line}). Neither
amplitudes, nor phases are precise but the frequencies should be well
described by this methodology. This kernel is used as initial guess
$\hat{K}_{PFS}^{0}$ to reach faster the convergence of the first
kernel estimation $\hat{K}_{PFS}^{1}$. Unfortunately, only Mars Express
orbit < 634 are usable for this estimation. It represents 310 orbits
out of 6255 orbits currently available i.e., less than 5\% of the
total current orbits.

\subsection{Laser line domain (5000 to 8330 cm$^{-1}$) \label{sub:Laser-line}}

Knowing that laser line is almost dirac shaped, we could first hypothesize
that the kernel $K_{PFS}$ can be directly measured in the laser line
domain. Unfortunately due to aliasing (laser line ghosts from left
and right are superposed), from $I_{PFS}(\sigma)$ it is not possible
to realize an inversion to estimate $K_{PFS}$ because there are twice
unknown variables in comparison to known variable. Under strong hypothesis,
it is possible to estimate an approximation of $\hat{K}_{PFS}^{approx}$
\citep{Shatalina_PFS_AOsubmitted}. 

Nevertheless, the recent analytical formulation of the mechanical
effects on the laser line, allows us to compute the exact effect of
the mechanical vibration on the laser line, knowing the vibration
kernel $K_{PFS}$ (see complex expression in \citet{Shatalina_PFS_AOsubmitted}).
After the estimation of $K_{PFS}$, one simple test of coherence would
be to compare the observed laser line ghost to the one predicted.

\subsection{Quality of the results}

We propose several criteria to estimate if the deconvolution is correct.

\subsubsection{Distance between real and simulated PFS spectra}

Because the only ground truth we could have is the real PFS spectra
$I_{PFS}$, it should be as close as possible to the final simulated
PFS spectra $\hat{I}_{PFS}^{final}=\hat{I}_{Mars}^{final}\star\hat{K}_{PFS}^{final}$.
We use the Root Mean Square distance (RMS) of $\hat{I}_{PFS}^{final}-I_{PFS}$.
In this way, we evaluate at the same time the correctness of the estimated
Mars spectra $\hat{I}_{Mars}^{final}$ and the instrument model $\hat{K}_{PFS}^{final}$.

\subsubsection{Ghost removal in the signal domain (0 to 5000 cm$^{-1}$)}

In the 1 to 1530 cm$^{-1}$ wavenumber domain, no signal is expected
due to the very low signal to noise ratio but only the ghosts are
present in the raw spectra. Thus, one simple criterion to estimate
the efficiency of the correction is to measure the energy in this
domain.

\subsubsection{Ghosts in the laser line domain (5000 to 8330 cm$^{-1}$)}

The laser line modulated $\hat{I}_{LM}^{final}$ through filter, aliasing
and vibrations effects can be computed from the estimated kernel $\hat{K}_{PFS}^{final}$
using the exact formulation of \citealp{Shatalina_PFS_AOsubmitted}.
To check the quality of the results, we evaluate the distance between
the actually measured signal and the predicted laser line modulated
with its ghosts.

\subsubsection{Distance to the approximated kernel}

The estimation of the kernel $\hat{K}_{PFS}^{approx}$ from the laser
line domain can be done. Neither amplitudes, nor phases are precise
but the frequencies should well described by this methodology. The
distance between $\hat{K}_{PFS}^{approx}$ and $\hat{K}_{PFS}^{final}$
is also a criteria of good results.

\subsubsection{Comparison with vibration frequencies from MEx telemetry and technical
specification}

Several sources of vibrations are present in the MEx platform, mainly
reaction wheels, Inertia Measurement Unit (IMU) dithering. PFS eigenmodes
can also be exited and are considered as ``source'' of vibrations.
Since, these vibrations are not unique onboard MEx (cryocooler, other
instruments, ...) and the uncertainties on these vibrations frequencies
are not known, it is not possible to have a supervised approach. One
also have to note that all vibration frequencies may not be present
in a PFS spectrum, depending the coupling with PFS. Nevertheless,
a comparison between our blind estimation and the actual data is interesting.

From each vibration frequency $f_{d}$ (in Hz), the perturbation is
at wavenumber $\sigma=f_{d}/v_{m}$ \citep{Saggin_PFSvibration_AO2007,Shatalina_PFS_AOsubmitted},
with the pendulum speed $v_{m}=d_{zc}.f_{zc}$ where the zero-crossing
frequencies $f_{zc}$ is 2500 Hz and zero-crossing length $d_{zc}$
is 1.2 microns for typical PFS measurements at Mars \citep{Giuranna_calibrationPFS-SWC_PSS2005}.

\paragraph*{Reaction wheels}

Thanks to telemetry data from ESA, it is possible to estimate the
frequencies of reaction wheels for ORB0032, spectra No 106
at 56.7 Hz, 33.3 Hz, 40.6 Hz and 30.3 Hz. Uncertainties are unknown
and those frequencies of micro-vibrations are expected to change during
the mission but can be estimated from telemetry.

\paragraph*{IMU}

Astrium technical specification of MEx (MEX.MMT.HO.2379) states that
the IMU dithering onboard MEx are at 513.9 Hz, 564.3 Hz and 617.4
Hz. Uncertainties are unknown but those frequencies of micro-vibrations
are expected to be constant during the mission.

\paragraph*{PFS eigenmodes}

The PFS eigenmodes are around 135 Hz and 160 Hz. Uncertainties are
unknown but those frequencies of micro-vibrations are expected to
be constant during the mission.

\section{Results}

Due to the stochastic character of the ghosts and especially their
phase, few \% of the PFS spectra in the archive, randomly distributed,
present significant level of perturbations. In some lucky cases, the
ghosts are absent but typical spectra contains few ghosts \citep{Comolli_ProbPFS_PSS2010}.
We propose to illustrate our algorithm on the ORB0032, spectra No
106 of PFS, recorded in particularly high level of disturbances. This
spectra contains several obvious ghosts (as shown by the arrows in
fig. \ref{fig:Final-results}). 

We find that the optimum inversion is reached with a loop of N=2,
with special parameter for the first step due using the filtered initialization
($\lambda_{K}$=50) and then usual parameter ($\lambda_{Mars}$=0.001,
$\lambda_{K}$=1) using the unfiltered spectra .

\subsection{Mars spectra and kernel estimations obtained }

\begin{figure}
\includegraphics[clip,width=1\textwidth]{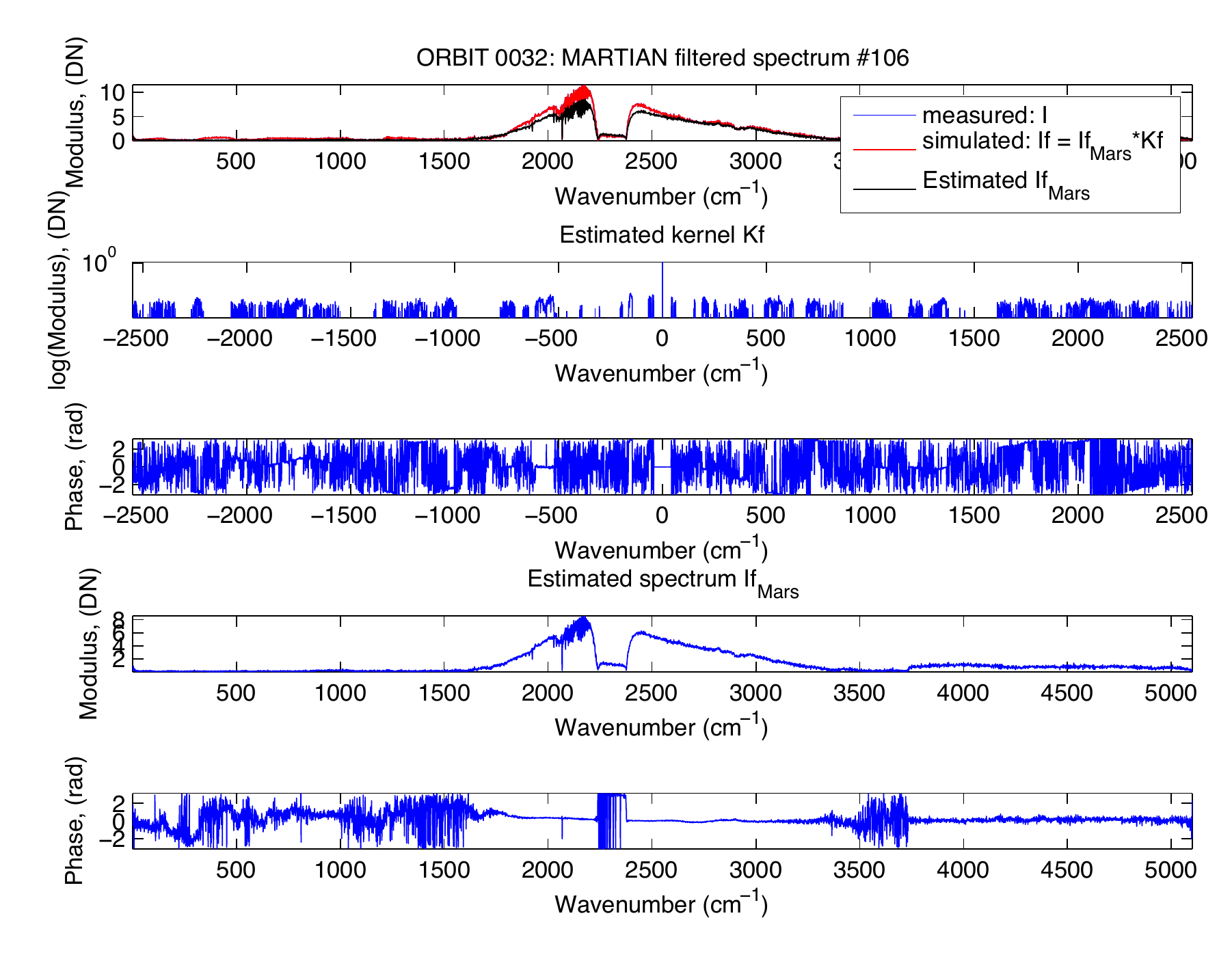}\center

\caption{Final results of the spectra ORB0032 \#106: at the top: modulus of
measured $I_{PFS}$ (blue) simulated $\hat{I}_{PFS}^{final}=\hat{I}_{Mars}^{final}\star\hat{K}_{PFS}^{final}$
PFS spectra (red) and the estimated Martian spectra $\hat{I}_{Mars}^{final}$
(black); Lack of fit between $I_{PFS}$ and $\hat{I}_{PFS}^{final}$
is 1.8 $10^{-5}$;in the middle: modulus (in log scale) and phase
of the final estimated kernel $\hat{K}_{PFS}^{final}$; at the bottom:
modulus and phase of the final estimated spectrum $\hat{I}_{Mars}^{final}$.\label{fig:FinalDeconvolutionSignal}}
\end{figure}

\begin{figure}
\includegraphics[clip,width=1\textwidth]{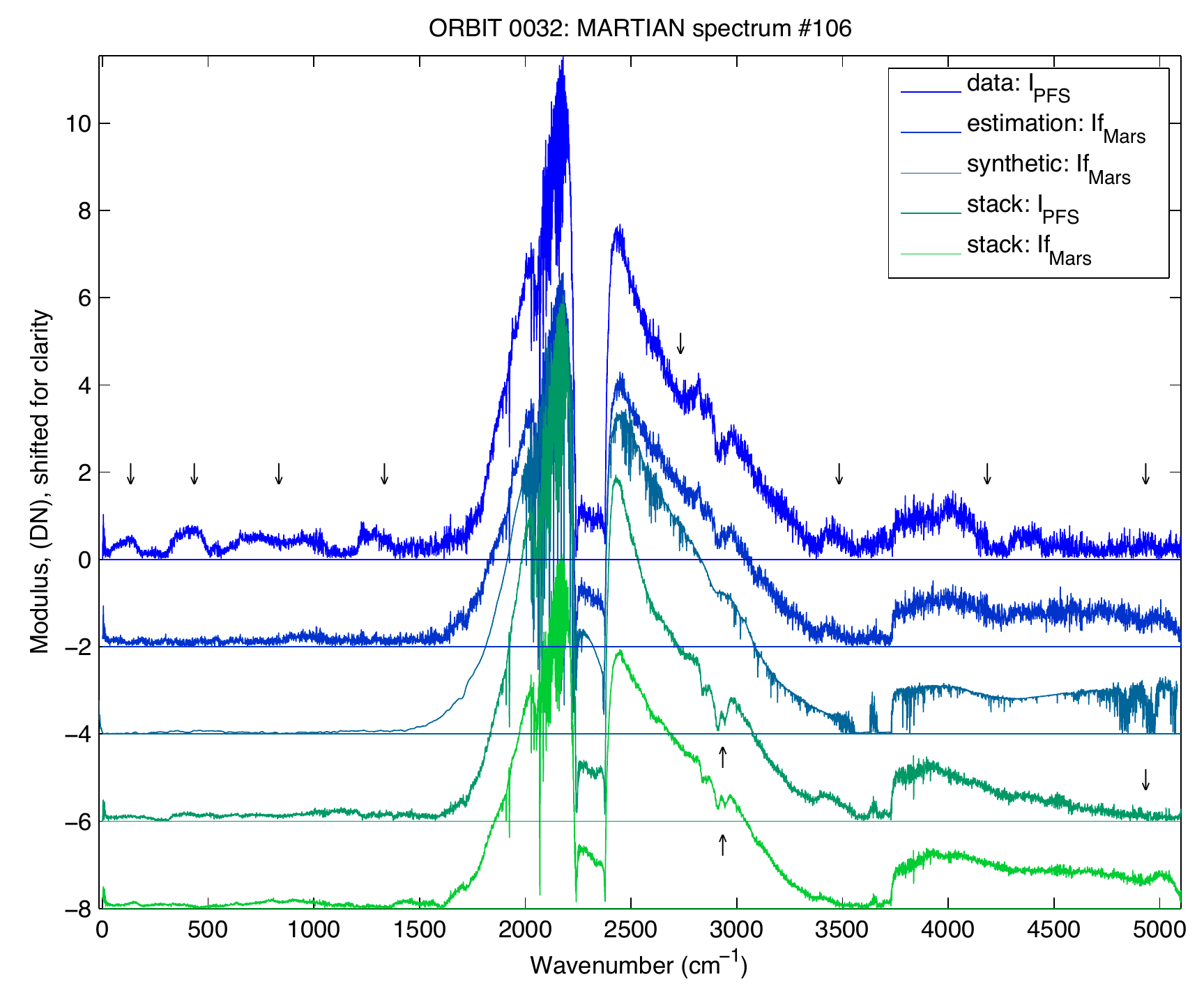}\center

\caption{Final results of the spectra ORB0032 \#106 as compared with stacking
and synthetic measurements, from top to bottom : (i) raw PFS measurements,
all arrows represents ghosts artifacts, (ii) estimated spectra from
our algorithm, (iii) synthetic measurement of PFS, (iv) stack of 11
PFS spectra, (v) stack of 11 estimated spectra from our algorithm.
The arrow at 2900 cm$^{-1}$ represents the mirror contamination by
hydrocarbons, the arrow at 4900 cm$^{-1}$ represents an artifact
of abnormal small signal, probably due to ghosts.\label{fig:Final-results}}
\end{figure}

For Mars spectra estimation, the final estimation of the signal $\hat{I}_{Mars}^{final}$
is presented in fig. \ref{fig:FinalDeconvolutionSignal} and \ref{fig:Final-results}.
This figure presents the raw spectra, our corrected spectra in comparison
with a synthetic spectra $\hat{I}_{Mars}^{0}$ (see section \ref{sub:Initial-guesses})
and also the stack of 20 spectra. Our correction clearly removes the
ghosts in the region at 1-1530 cm$^{-1}$, around 2700 cm$^{-1}$,
around 3450 cm$^{-1}$, around 4150 cm$^{-1}$ similar to the stacking
method. The artifact at 2900 cm$^{-1}$ persists, due to pollution
of hydrocarbons in the telescope \citep{Giuranna_calibrationPFS-SWC_PSS2005}.
In the 4000-5000 cm$^{-1}$ domain, our method improves the signal
in comparison to the stacking method and partly correct the artificial
decrease of the signal. The stacking clearly reduces the stochastic
noise, that is not removed with our correction. 

\begin{figure}
\includegraphics[clip,width=1\columnwidth]{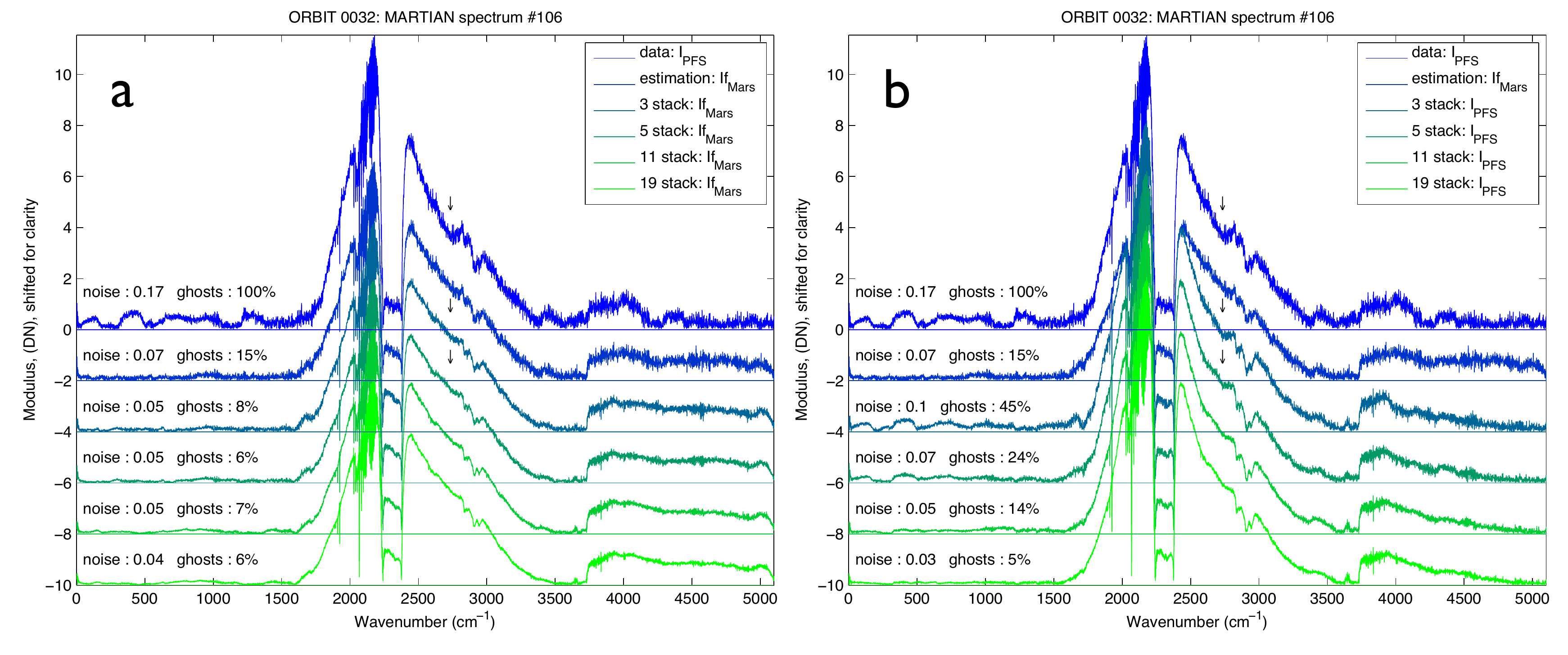}\center

\caption{Comparison of our correction versus the stacking method : (a) Stacking
of corrected spectra from our method (b) Stacking of PFS spectra.
Noise standard deviation from the 1-1530 cm$^{-1}$ are expressed
for all spectra. Fraction of energy due to ghosts, relative to the
raw PFS spectra in 1-1530 cm$^{-1}$ is also written for all spectra.
Arrows at 2700 cm$^{-1}$ represent significant difference in the
signal domain due to ghosts, that persists for stacking of at least
5 PFS spectra but well corrected by our method.\label{fig:Final-results-stack}}
\end{figure}

Figure \ref{fig:Final-results-stack} shows the evolution of the average
spectra, when stacking 3, 5, 11 and 19 spectra. The plots clearly
show that our method removes the ghosts contribution, already without
stacking. In contrary, the stacking methods require $\sim$10 spectra
to remove this effect. The signal to noise ratio at small scale, estimated
by the standard deviation in the 1-1530 cm$^{-1}$, is not significantly
changed between both methods.

The stack of $\sim$10 spectra correspond to $\sim$10 spots of around
7 km each, so that the spatial resolution can be improved by one order
of magnitude. In term of temporal resolution improvement, it depends
mainly on the location due to the very irregular observation density.

\subsection{Quality of the results}

As illustrated in Fig. \ref{fig:FinalDeconvolutionSignal}, the lack
of fit between the real PFS spectra $I_{PFS}$ and the simulated one
from our final guesses $\hat{I}_{PFS}^{final}=\hat{I}_{Mars}^{final}\star\hat{K}_{PFS}^{final}$
$ $is very small ($\sim10^{-5}$), showing that the solution is compatible
with the observation.

Our strategy is efficient to remove the norm in the 1 to 1530 cm$^{-1}$
domain by a factor of $\sim2$ (the RMS is 0.0093 for the raw spectra
and to 0.0042 for the corrected spectra). The only signal left in
$\hat{I}_{Mars}^{final}$ seems to be very small scale random, as
expected (see Fig. \ref{fig:FinalDeconvolutionSignal}). The theoretical
value of noise standard deviation is about 0.1 using the estimated
Signal to Noise ratio of about 100 in the 2000-2400 cm$^{-1}$ \citet{Giuranna_calibrationPFS-SWC_PSS2005}.
The estimated noise standard deviation of corrected spectra is in
agreement with this value (see fig. \ref{fig:Final-results-stack}).
In order to estimate the efficiency of ghost removal, we measure the
spectral energy in the 1 to 1530 cm$^{-1}$ domain at scale larger
than 50 cm$^{-1}$ for corrected spectra in comparison to the measured
spectra, assuming that the large scale features are only due to ghosts.
We found that our correction for one single spectra remove 85\% of
the ghost energy, which is equivalent to the effect of stacking of
11 spectra.

\begin{figure}
\includegraphics[clip,width=1\textwidth]{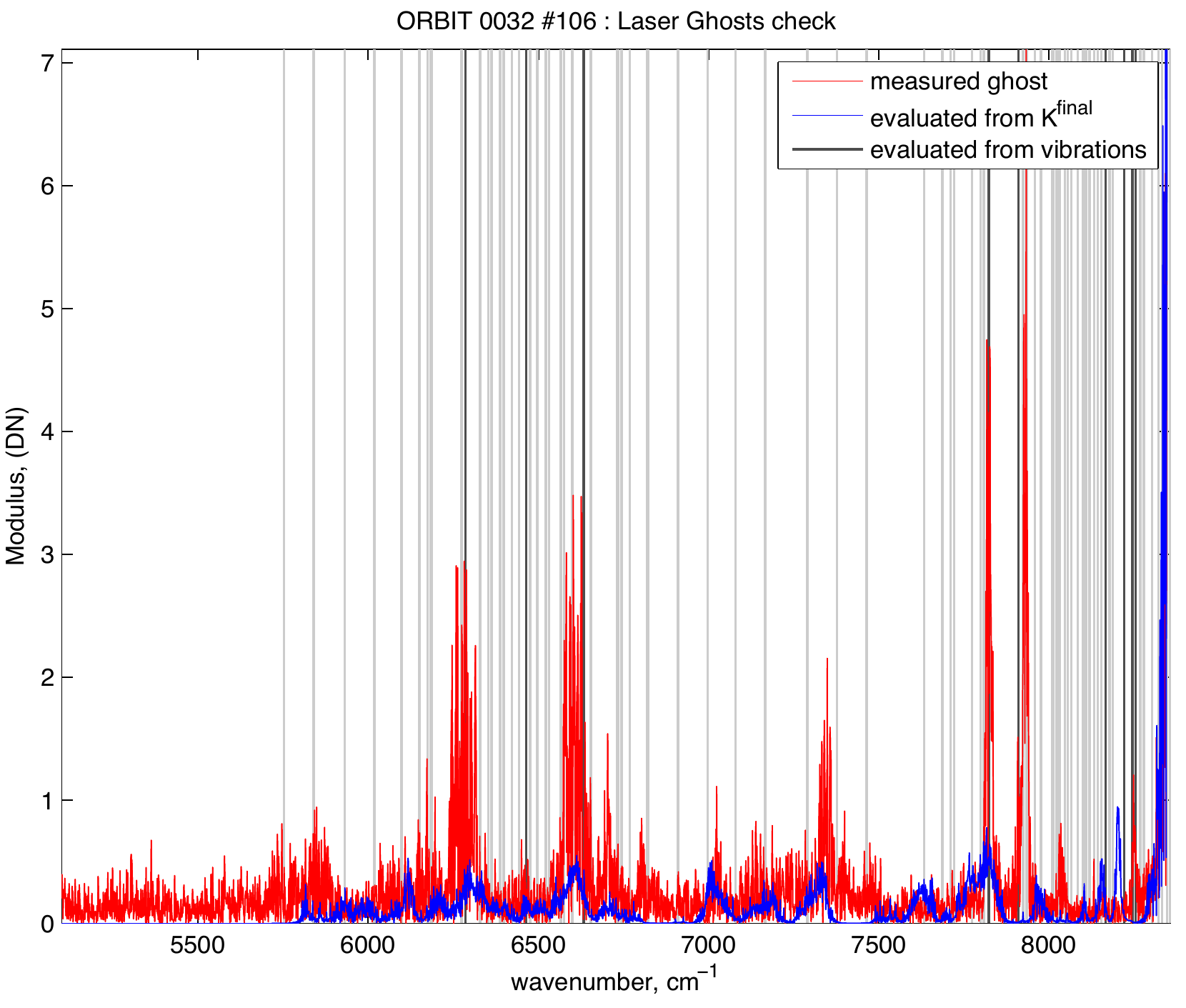}\center

\caption{Modulus of the simulated laser line modulated $\hat{I}_{LM}^{final}$
through filter, aliasing and vibration kernel $\hat{K}_{PFS}^{final}$
(blue) and the observation $I_{LM}$ (red). The lack of fit is 0.0270\label{fig:LaserLineGhosts}}
\end{figure}

The laser line modulated $\hat{I}_{LM}^{final}$ through filter, aliasing
and vibration kernel $\hat{K}_{PFS}^{final}$ are compatible with
the observation $I_{LM}$ (see Fig. \ref{fig:LaserLineGhosts}). The
four mains peaks are estimated and also some smaller peaks. The distance
is relatively small ($\sim0.013$). 

\begin{figure}
\includegraphics[clip,width=1\textwidth]{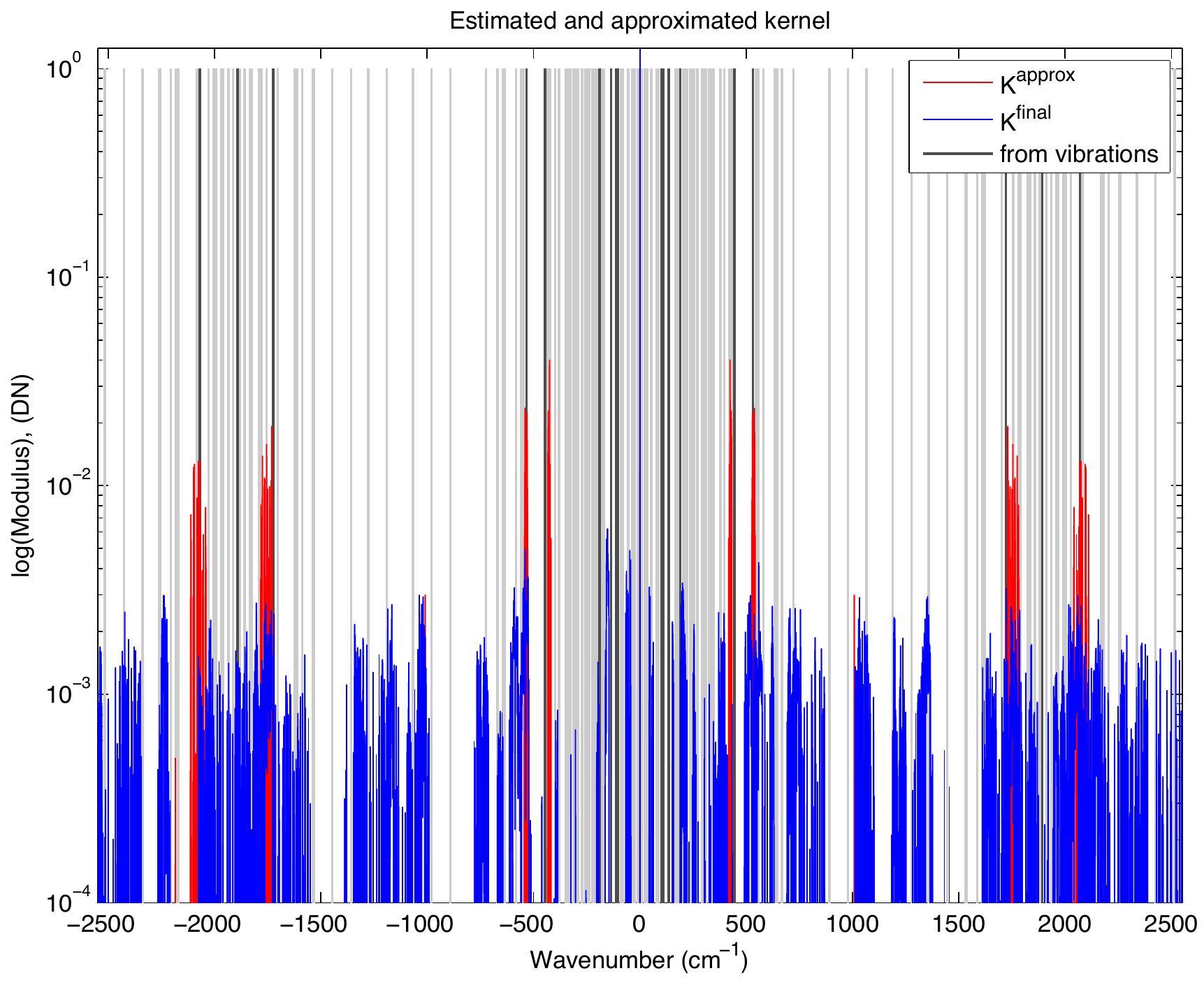}\center

\caption{Modulus in log scale of the vibration kernel $\hat{K}_{PFS}^{final}$
(blue line), the approximated kernel $\hat{K}_{PFS}^{approx}$ (red
line)and the reaction wheels vibration (dark grey line), and the combination
of reaction wheels (light grey line). The lack of fit is 3.2 $10^{-4}$.
\label{fig:DistanceEstimatedKernel}}
\end{figure}

The final kernel estimation $\hat{K}_{PFS}^{final}$ is close to the
initial kernel guess $\hat{K}_{PFS}^{0}=\hat{K}_{PFS}^{approx}$ with
a distance $\sim10^{-5}$. As illustrated in Fig. \ref{fig:DistanceEstimatedKernel},
the main vibration frequencies estimated in $\hat{K}_{PFS}^{approx}$
are present in $\hat{K}_{PFS}^{final}$. The estimation of $\hat{K}_{PFS}^{approx}$
has been done under strong approximation. Especially the unconstrained
amplitude may explain the differences. Also $\hat{K}_{PFS}^{final}$
presents a smooth signal due to the high frequencies filtering. Other
methods without sparsity regularization doesn't succeed to get such
a sparse kernel although we believe that the kernel is sparse due
to limited vibrations in the mechanical environment of PFS onboard
MEx (eigenmode of PFS, reaction wheels frequencies, inertia measurement
unit dithering frequencies). 

\section{Discussion and conclusion}

We described the approximated direct problem and an algorithm able
to correct for the mechanical vibration of the PFS instrument. For
the first time, we show that it is possible to reduce significantly
the ghosts from the observed signal from 3-5 \% of the total energy
to 0.4-0.7 \%. We show that our estimation is coherent using three
quantities: ghosts in the signal domain, laser line ghosts, distance
to approximated kernel. Thus the global shape of PFS SWC spectra can
be corrected with our algorithm, allowing to better estimate temperature,
and thermal profile on each PFS measurement, improving the few \%
of spectra with high $\chi^{2}$ that could not be processed with
current calibration \citep{Grassi_Methodsanalysisdata_PaSS2005}.
Also, our correction may avoid the continuum removal step in the minor
species retrieval \citep{Sindoni_WaterPFS_PSS2011}. When the signal
to noise ratio is high enough, our correction will also reduce the
stacking procedure.

In the future, we would like to propose an algorithm to correct the
complete archive that would require: efficient algorithm, timesaving
implementation, and fully automatic procedure. Also, new correction
procedure must be developed to treat the whole orbits currently available
(6405 at the date of writing).

In order to correct any shaked FTS, semi-blind deconvolution is possible,
knowing $\hat{I}_{Mars}^{0}$ (but without knowing $\hat{K}_{PFS}^{approx}$
from the \textquotedblleft{}laser line domain\textquotedblright{})
so that the \textquotedblleft{}signal domain\textquotedblright{} only
is required. Thus, any techniques of optical path measurement (laser
line, mechanical, etc\ldots{}) can be corrected with our technique.
Nevertheless, the independent estimation of the kernel $\hat{K}_{PFS}^{approx}$
significantly improve the convergence of the algorithm. The only limitation
to apply this method on other instruments is about the convolution
equation. Convolution is true if the signal from the planet $I_{Mars}$
has a significantly reduced wavenumber domain (as stated in eq. 1-3).

\subsubsection*{Acknowledgement}

We thank Ali Mohammad-Djafari for fruitful discussions. We acknowledge
support from the ``Institut National des Sciences de l'Univers''
(INSU), the ``Centre National de la Recherche Scientifique''
(CNRS) and ``Centre National d'Etude Spatiale''
(CNES) and through the ``Programme National de Plan{\'e}tologie''.
We also thank ``European Space Agency'' (ESA) for providing the
reaction wheels speed of MEx.

\bibliographystyle{elsarticle-harv}

\end{document}